# Artifact side peaks in saturated absorption spectroscopy caused by inappropriate lock-in amplification

Leo Matsuoka[1]*

[1]*Faculty of Engineering, Hiroshima Institute of Technology, Hiroshima, Hiroshima 731-5193, Japan*

E-mail: r.matsuoka.65@cc.it-hiroshima.ac.jp

Saturated absorption spectroscopy is a powerful technique that provides valuable information about gas-phase atomic targets, including not only the precise positions of transitions but also the effects of velocity-changing collisions, which manifest as changes in the shape of the peaks. However, using a lock-in amplifier inappropriately can generate artifact side peaks, potentially leading to a misinterpretation of results. To investigate the emergence of these artifacts, we conducted numerical calculations based on the basic Lambert-Beer's law. Our model reproduced the artifact side peaks and revealed that these artifacts become more significant as the transmittance of the probe laser decreases.

Saturated absorption spectroscopy is a fundamental Doppler-free spectroscopic technique that has been widely used for detecting hyperfine structures. [1,2] This simple method involves taking the difference in the absorption spectra with and without the pump laser incident in opposite directions. It has been utilized in student experiments and is increasingly applied in plasma research as a high-precision wavelength standard, [3] for electric field measurement, [4] for electron density measurement, [5] and to observe the effect of velocity-changing collisions. [6] The shapes of saturated absorption spectra are actively studied in theoretical research, and various computational models of spectra under different conditions, including the influence of polarization, are being developed.[7-9] Saturated absorption spectra also play a fundamental role in advanced Doppler-free spectroscopy techniques, such as Modulation Transfer Spectroscopy. [10-12]

Saturated absorption spectroscopy relies on the saturation of absorption by the zero-velocity component along the laser axis in the gas-phase target atoms induced by the pump laser incident in opposite directions. However, this method assumes that the excited atoms do not undergo velocity-changing collisions (VCC) during excitation by the pump laser. When pressure and temperature increase, VCC can occur, resulting in various changes in the saturated absorption spectrum depending on collision rates in the upper and lower levels, as well as the deactivation rate, which includes spontaneous emission. VCC has been studied both experimentally [13-18] and theoretically. [19,20] The analysis of VCC can provide valuable information about the target. Although VCC research dates back to the previous century, there is still potential for further theoretical and experimental exploration.

The lock-in amplifier is a widely employed instrument in saturated absorption spectroscopy. In this method, the peaks appear as small variations on a prominent absorption signal. As a standard practice, the intensity of the pump laser is modulated when using a lock-in amplifier, amplifying only the changes in the absorption signal that result from this modulation. This approach allows for the detection of very small peaks or peaks that undergo gradual changes over time. Regarding VCC, another effective application of the lock-in amplifier exists. When VCC broadens the saturated absorption spectrum, the peaks can be sharpened with modulation frequencies faster than the VCC frequency. [21,22] Conversely, this phenomenon offers the potential to quantify the VCC frequency of the target using frequency modulation of the pump laser.

However, caution is required when using a lock-in amplifier in saturated absorption spectroscopy. According to Lambert-Beer's law, the target density spectrum is determined by taking the logarithm of the reciprocal of the transmittance. To calculate the density spectrum of the saturated absorption spectrum, we need to compute the logarithm of the ratio of the transmittance with and without the pump laser. When the lock-in amplifier is used without logarithmic signal preprocessing, it simply amplifies the difference between them. The peak shape observed by the lock-in amplifier without signal preprocessing does not accurately represent the true spectrum. Therefore, caution is necessary, especially when examining the spectral shape beyond the positions of the peaks.

In this study, we employed numerical calculations based on Lambert-Beer's law to examine artifact side peaks that result from the improper use of a lock-in amplifier in saturated absorption spectroscopy. These artifacts hold the potential to lead to a misinterpretation of the effects, such as alterations in spectra due to factors like VCC. As far as our knowledge extends, previous literature has not explicitly addressed artifact side peaks, despite their actual observation.

Here, we describe the computational model used in this study. The intensity of light transmitted through a target with absorption attenuates based on the Lambert-Beer's law, as described by

$$I(\lambda) = I_0 \exp\left(-\varepsilon L C(\lambda)\right), \tag{1}$$

where $I(\lambda)$ is the transmitted intensity of light for the relative wavelength $\lambda$, $I_0$ is the initial intensity of the light, $\varepsilon$ is the proportionality constant based on the absorption intensity of atoms, $L$ is the optical path length, and $C(\lambda)$ is the relative density of the target that interacts with the light of $\lambda$. In this study, we consider $C(\lambda)$ as a single-peak function that reaches its maximum value at $\lambda = 0$. The absolute value of the atomic density is included in $\varepsilon L$.

Figure 1 depicts the simplified experimental setup used in this study. In saturated absorption spectroscopy, a single laser emitting at a particular wavelength is divided into two beams and directed towards the target from opposite directions. For simplicity, we assume that the laser intensity $I_0$ is uniform across all wavelengths, and we neglect the attenuation resulting from the optical path of the pump laser, focusing solely on the attenuation of the probe laser. Additionally, we do not consider any absorption saturation effects caused solely by the pump laser or solely by the probe laser.

We define a model density spectrum of the saturated absorption spectroscopy with pump laser. This spectrum can be expressed as the difference between a Gaussian spectrum with a Doppler width, and a Lorentzian spectrum which represents a density depression when saturation absorption occurs. Although the depression due to the pump laser occurs at positions other than the center of the spectrum, it is only noticeable near the center where the pump and probe lasers interact with atoms that have the same velocities (nearly zero axial velocity). Therefore, we can fix the Lorentzian component at the center. The Gaussian component $C_{\mathrm{G}}(\lambda)$, the Lorentzian component $C_{\mathrm{L}}(\lambda)$, and the model spectrum $C_{\mathrm{P}}(\lambda)$ are defined as follows,

$$C_{\mathrm{G}}(\lambda) = \exp(-\lambda^2/0.02)\,, \tag{2}$$

$$C_{\mathrm{L}}(\lambda) = 0.1\left(\frac{\lambda^2}{0.00005}+1\right)^{-1}, \tag{3}$$

$$C_{\mathrm{P}}(\lambda) = C_{\mathrm{G}}(\lambda) - C_{\mathrm{L}}(\lambda). \tag{4}$$

Here, the height of the Lorentzian component is set to 0.1 times the height of the Gaussian component, and the width was set to match typical experimental results. The model spectrum and each component are shown in Fig. 2(a). In experiments, the objective of the saturated absorption spectroscopy is to derive $C_{\mathrm{L}}(\lambda)$ by $C_{\mathrm{P}}(\lambda)$ and $C_{\mathrm{G}}(\lambda)$, measured with and without the injection of the pump laser.

We now express the transmittance based on Eq. (1), where we define $-\varepsilon L = \ln T$, with $T$ being the transmittance at the peak of the absorption spectrum. When $C_{\mathrm{P,G}}(\lambda) = 1$, the transmittance is equal to $T$. The experimental signals with and without the pump laser are expressed by the following equations:

$$S_{\mathrm{P}}(T,\lambda) = I_0 \exp(C_{\mathrm{P}}(\lambda)\ln T)\ , \tag{5}$$

$$S_{\mathrm{G}}(T,\lambda) = I_0 \exp(C_{\mathrm{G}}(\lambda)\ln T)\ . \tag{6}$$

In Figs. 2(b) and (c), we show the transmittance spectra $S_{\mathrm{P}}(T,\lambda)/I_0$ and $S_{\mathrm{G}}(T,\lambda)/I_0$ for $T$ = 0.5 and $T$ = 0.005 as an example, respectively. When the pump laser is present, a depression in absorption occurs around the peak position. However, in terms of appearance, it can be observed that there is not much difference depending on $T$ values.

Our goal is to determine $C_{\mathrm{L}}(\lambda)$ from the measurement, and this is achieved by the following equations:

$$\begin{aligned} C_{\mathrm{L}}(\lambda) &= C_{\mathrm{G}}(\lambda) - C_{\mathrm{P}}(\lambda) \\ &= \frac{1}{\ln T}\left(\ln\frac{S_{\mathrm{G}}(T,\lambda)}{I_0} - \ln\frac{S_{\mathrm{P}}(T,\lambda)}{I_0}\right) \\ &= \frac{1}{\ln T}\ln\frac{S_{\mathrm{G}}(T,\lambda)}{S_{\mathrm{P}}(T,\lambda)}\ \big(=: S_{\mathrm{L}}(T,\lambda)\big)\ . \end{aligned} \tag{7}$$

On the other hand, when using a lock-in amplifier by modulating the intensity of the pump laser, the signal is proportional to the difference between two spectra:

$$D_{\mathrm{L}}(T,\lambda) \propto S_{\mathrm{P}}(T,\lambda) - S_{\mathrm{G}}(T,\lambda)\,, \tag{8}$$

where the absolute magnitude depends on the settings of the lock-in amplifier.

Figure 3 compares $S_\mathrm{L}(T,\lambda)$ and $D_\mathrm{L}(T,\lambda)$ at various $T$ values. Although the absolute values of $D_\mathrm{L}(T,\lambda)$ differ, appropriate coefficients have been applied to scale them to match their maximum values at $\lambda = 0$. $S_\mathrm{L}(T,\lambda)$ remains constant regardless of the value of $T$, while $D_\mathrm{L}(T,\lambda)$ exhibits increasing side peaks as the transmission decreases. However, the central peak remains relatively unchanged. These side peaks are artifact peaks that appear in the tail region and can be mistaken for the VCC peaks or so.

In a certain previous paper, [18] these artifact side peaks were observed, and there was a potential misunderstanding that they were due to some effect like VCC. In that experiment, saturated absorption spectroscopy was conducted using a uranium hollow cathode lamp with a lock-in amplifier under low pressure and high current conditions, producing signals resembling the artifact peaks observed in this study. While the researchers speculated that these artifacts were due to the level splitting caused by high voltage, it is more reasonable to consider them as artifacts in this study. Firstly, the shapes of the peaks are similar. Secondly, under high-current conditions, the plasma density increases, resulting in low transmittance of the probe laser. Additionally, it is reasonable to observe these artifacts only under conditions where VCC does not strongly affect the shape of the peaks.

Under low transmission conditions, these peaks become more prominent because the apparent spectrum deviates significantly from linearity due to Lambert-Beer's law. To avoid the emergence of these artifact peaks while still using the lock-in amplifier, you can employ three strategies. Firstly, you can apply logarithmic preprocessing using a balanced detector. Secondly, you can achieve a lower target density to increase the transmittance of the probe laser. Thirdly, while increasing $I_0$ has no effect based on our calculation model, one can simply increase $I_0$. In real experiments, due to the saturation effect, the transmittance intensity of the probe laser deviates from the Lambert-Beer’s law and approaches linear decrease given by,

$$I(\lambda) = I_0 - \varepsilon_\mathrm{s} L C(\lambda)\,, \tag{9}$$

where $\varepsilon_\mathrm{s}$ represents the saturation absorption intensity per unit length. In this region, the simple difference obtained by the lock-in amplifier starts to appear as representing the correct saturated absorption spectrum. However, in saturated absorption spectroscopy with multiple peaks, such as cesium, the relative heights of the saturated absorption spectrum peaks can vary depending on the intensity of the probe laser. Therefore, merely increasing $I_0$ does not guarantee the same spectrum.

Saturated absorption spectroscopy has been advanced by lock-in amplifiers, which not only enhance the signal clarity but also make active measurement of collision frequencies possible. However, it has been demonstrated that simply turning the pump laser on and off may cause significant deviation from the actual spectrum.

In conclusion, this study demonstrated the possibility of artifact peaks appearing in saturated absorption spectroscopy using a lock-in amplifier. While saturated absorption spectroscopy is a widely used technique that appears in textbooks, many aspects of phenomena that occur when conditions deviate from the ideal are still not fully understood. Some aspects have been inherited as implicit knowledge, but the sharing of implicit knowledge regarding spectrum interpretation is still necessary.

**Acknowledgments**

This work was supported by JSPS KAKENHI Grant Number JP21K04947.

## References

1) E.C. Jung, D.-Y. Jeong, K. Song, J. Lee, Opt. Commun. **141**, 83-90 (1997).
2) H. Liu, W. Yuan, F. Cheng, Z. Wang, Z. Xu, K. Deng, and Z. Lu, J. Phys. B: At. Mol. Opt. Phys. **51**, 225002 (2018).
3) M. Aramaki, K. Ogiwara, S. Etoh, S. Yoshimura, and M. Y. Tanaka, Rev. Sci. Instrum. **80**, 053505 (2009).
4) S. Nishiyama, H. Nakano, M. Goto, and K. Sasaki, J. Phys. D: Appl. Phys. **50**, 234003 (2017).
5) A. Zafar, E. Martin, and S. Shannon, J. Quant. Spectrosc. Radiat. Transfer **230**, 48-55 (2019).
6) N. Nishiya and L. Matsuoka, IEEE Transactions on Plasma Science **47**, 1129-1133 (2019).
7) G. Moon and H.-R. Noh, J. Korean Phys. Soc. **50**, 1037-1043 (2007).
8) G. Moon and H.-R. Noh, J. Opt. Soc. Am. B **25**, 701-711 (2008).
9) H.-R. Noh, J. Korean Phys. Soc. **57**, 1381 - 1386 (2010).
10) D. J. McCarron, S. A. King and S. L. Cornish, Meas. Sci. Technol. **19**, 105601 (2008).
11) S. E. Park and H.-R. Noh, Opt. Express **21**, 14066-14073 (2013).
12) H.-R. Noh and S. E. Park, Opt. Commun. **336**, 173-176 (2015).
13) P. W. Smith and R. Hänsch, Phys. Rev. Lett. **26**, 740-743 (1971).
14) C. Brechignac, R. Vetter, and P. R. Berman, Phys. Rev. A **17**, 1609-1613 (1978).
15) P. Cahuzac, E. Marie, O. Robaux, R. Vetter, and P. R. Berman, P Cahuzac, J. Phys. B: Atom. Mol. Phys. **11**, 645-651 (1978).
16) P. F. Liao, J. E. Bjorkholm, and P. R. Berman, Phys. Rev. A **21**, 1927-1938 (1980).
17) C. G. Aminoff, J. Javanainen, and M. Kaivola, Phys. Rev. A **28**, 722-737 (1983).
18) G.K. Bhowmick, B.N. Jagatap, S.A. Ahmad, and V.B. Kartha, Spectrochim. Acta A: Mol. Spectrosc. **48**, 1539-1546 (1992).
19) P. R. Berman, Phys. Rev. A **13**, 2191-2211 (1976).
20) J. Tenenbaum, E. Miron, S. Lavi, J. Liran, M. Strauss, J. Oreg, and G. Erez, J. Phys. B: At. Mol. Phys. **16**, 4543-4553 (1983).
21) H.-D. Kronfeldt, G. Klemz, and D. Ashkenasi, Opt. Commun. **110**, 549-554 (1994).

22) D.-Y. Jeong, K. Song, J. Han, J. Lee, and B. K. Lee, Opt. Commun. **153**, 226-230 (1998).

## Figure Captions

**Fig. 1.** Conceptual diagram of the saturated absorption spectroscopy addressed in this study.

**Fig. 2.** (a) Model spectrum and each component used in the calculation (b) Calculation results of transmittance spectra at $T$ = 0.5 (c) Calculation results of transmittance spectra at $T$ = 0.005.

**Fig. 3.** The saturated absorption spectra by assuming inappropriate lock-in amplification $D_{\mathrm{L}}(T, \lambda)$ at each $T$ value, and the saturated absorption spectra calculated by properly considering Lambert-Beer's law $S_{\mathrm{L}}(T, \lambda)$. The inset shows magnified view of the spectra.

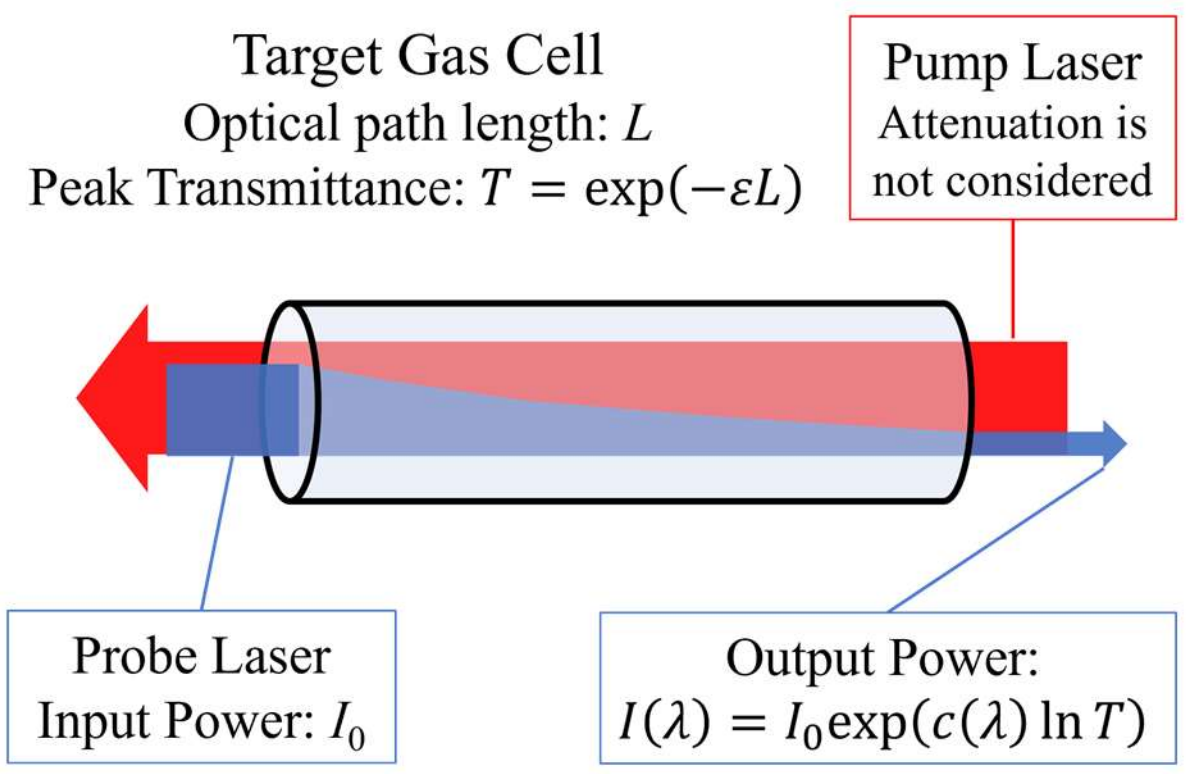
Target Gas Cell
Optical path length: $L$
Peak Transmittance: $T = \exp(-\varepsilon L)$
Pump Laser
Attenuation is not considered
Probe Laser
Input Power: $I_0$
Output Power:
$I(\lambda) = I_0 \exp(c(\lambda) \ln T)$

Fig.1.

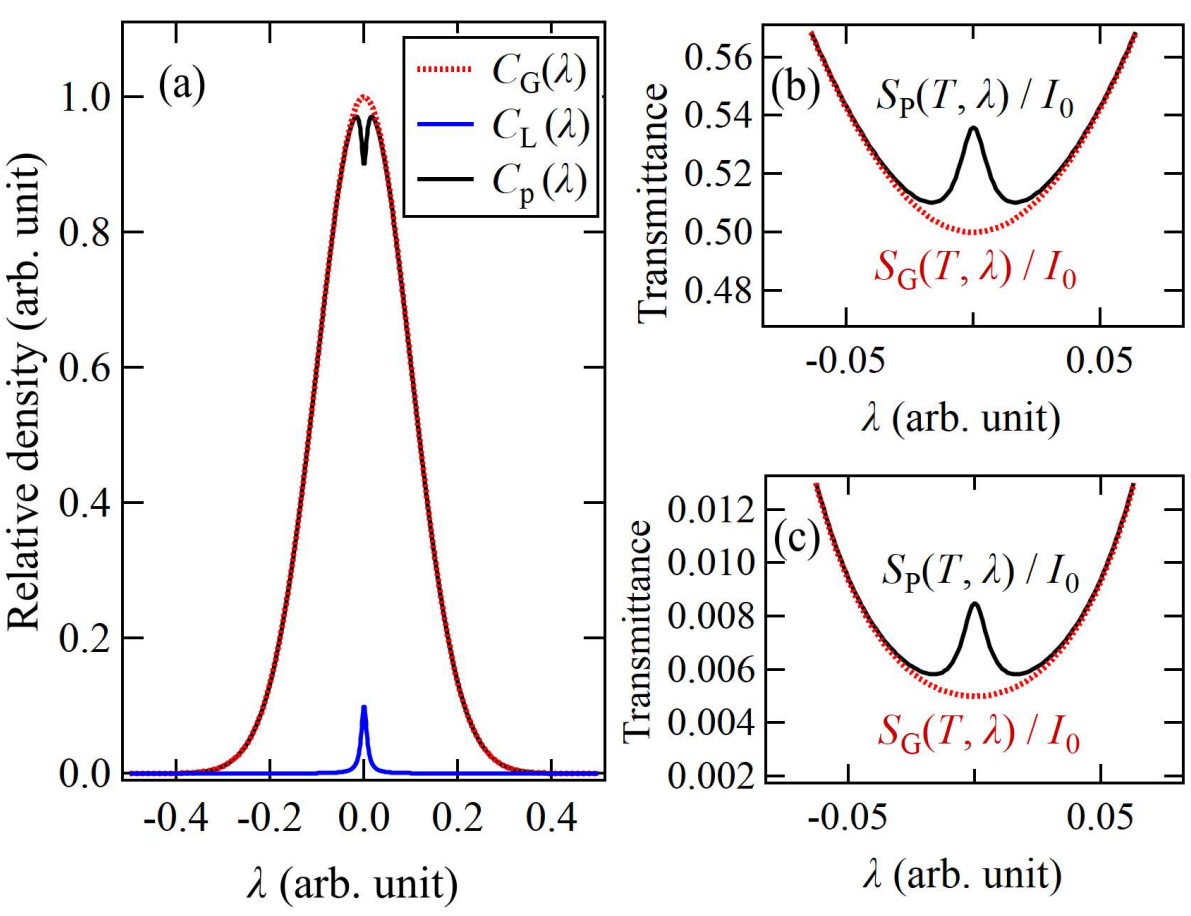
(a)
$C_G(\lambda)$
$C_L(\lambda)$
$C_p(\lambda)$
Relative density (arb. unit)
1.0
0.8
0.6
0.4
0.2
0.0
-0.4
-0.2
0.0
0.2
0.4
$\lambda$ (arb. unit)
(b)
$S_P(T, \lambda) / I_0$
$S_G(T, \lambda) / I_0$
Transmittance
0.56
0.54
0.52
0.50
0.48
-0.05
0.05
$\lambda$ (arb. unit)
(c)
$S_P(T, \lambda) / I_0$
$S_G(T, \lambda) / I_0$
Transmittance
0.012
0.010
0.008
0.006
0.004
0.002
-0.05
0.05
$\lambda$ (arb. unit)

Fig. 2.

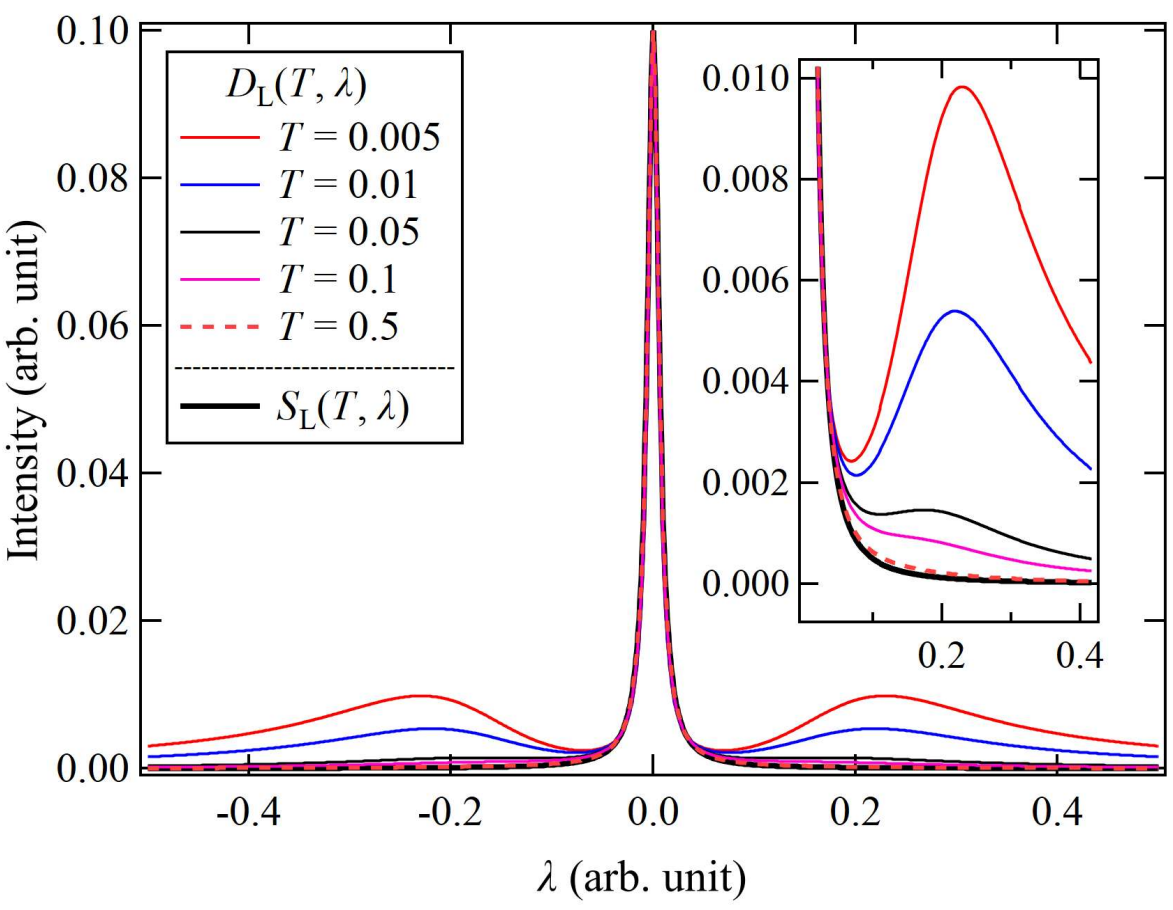

$D_L(T, \lambda)$
$T = 0.005$
$T = 0.01$
$T = 0.05$
$T = 0.1$
$T = 0.5$
$S_L(T, \lambda)$
Intensity (arb. unit)
$\lambda$ (arb. unit)

Fig. 3.